\documentclass[oldversion,paper]{aa} % for the abstract without structuration
\usepackage{natbib}

\usepackage{subfigure}
\usepackage{graphicx}
\usepackage{txfonts}
\usepackage{epstopdf}
\begin{document}

   \title{Multi-band transit observations of the TrES-2b exoplanet}
   %\title{Multi-band photometric indication for inclination changes of the TrES-2 exoplanet}

   \author{D. Mislis$^{1}$, S. Schr\"{o}ter$^{1}$, J.H.M.M. Schmitt$^{1}$, O. Cordes$^{2}$,  \and  K. Reif$^{2}$ }
   \authorrunning{D. Mislis, S. Schr\"{o}ter, J.H.M.M. Schmitt, O. Cordes, \and K. Reif}

\institute{$^{1}$Hamburger Sternwarte, Gojenbergsweg 112, D-21029 Hamburg \\
           $^{2}$Argelander-Institut f\"{u}r Astronomie, Auf dem H\"{u}gel 71, D-53121 Bonn\\
              \email{mdimitri@hs.uni-hamburg.de}
             }

\date{Accepted 2009 December 14}

\abstract{

We present a new data set of transit observations of the TrES-2b exoplanet taken in spring 2009,  using the 1.2m Oskar-L\"{u}hning telescope (OLT) of Hamburg Observatory and the 2.2m telescope at 
Calar Alto Observatory using BUSCA (Bonn University Simultaneous CAmera). Both the new OLT data, taken with the same instrumental setup as our data taken in 2008, as well as the simultaneously recorded multicolor BUSCA data confirm the low 
inclination values reported previously, and in fact suggest that the TrES-2b exoplanet has already passed the first inclination threshold $(i_{min,1} = 83.417^{\circ})$ and is not eclipsing the full stellar 
surface any longer. Using the multi-band BUSCA data we demonstrate that the multicolor light curves can be consistently fitted with a given set of limb darkening coefficients without the need to 
adjust these coefficients, and further, we can demonstrate that wavelength dependent stellar radius changes must be small as expected from theory. Our new observations provide further evidence for a 
change of the orbit inclination of the transiting extrasolar planet TrES-2b reported previously. We examine in detail possible causes 
for this inclination change and argue that the observed change should be interpreted as nodal regression.  While the assumption of an oblate host star requires an unreasonably large second harmonic coefficient, 
the existence of a third body in the form of an additional planet would provide a very natural explanation for the observed secular orbit change. Given the lack of clearly observed short-term variations of transit timing 
and our observed secular nodal regression rate, we predict a period between approximately 50 and 100 days for a putative 
perturbing planet of Jovian mass. Such an object should be detectable with present-day radial velocity (RV) 
techniques, but would escape detection through transit timing variations.}

\keywords{Stars : planetary systems -- Techniques : photometry}

\maketitle

\section{Introduction}

As of now more than 400 extrasolar planets have been detected around solar-like stars.
In quite a few cases several planets have been detected to orbit a given star,
demonstrating the existence of extrasolar planet systems in analogy to our solar system.
%,which harbors -- according to IAU definitions -- a total of eight planets.
Just as the planets in our solar
system interact gravitationally, the same must apply to extrasolar planet systems.
Gravitational interactions are important for the understanding of the long-term dynamical stability
of planetary systems.  The solar system has been around for more than four billion years, and
the understanding of its dynamical stability over that period of time is still a challenge \cite{1994A&A...282..663S}. In analogy, extrasolar planet systems must be dynamically stable over similarly long time scales, 
and most stability studies of extrasolar planet systems have been directed towards an understanding of exactly those long time scales. Less attention has been paid to secular and short-term perturbations of the orbit, 
since such effects are quite difficult to detect observationally. \cite{2002ApJ...564...60M} gives a detailed discussion on what secular effects might be derivable from extrasolar planet transits; 
in spectroscopic binaries the orbit inclination $i$ can only be deduced in conjunction with the companion mass, and therefore the detection of an orbit inclination change is virtually impossible.
Short-term Transit timing variations (TTVs) have been studied by a number of authors (\cite{2005Sci...307.1288H, 2005MNRAS.359..567A}), however, a detection of such effects has remained
elusive so far. 
In principle, a detection of orbit change would be extremely interesting since it would open up entirely new diagnostic possibilities of the masses and orbit geometries of these systems; also, in 
analogy to the discovery of Neptune, new planets could in fact be indirectly detected.

The lightcurve of a transiting extrasolar planet with known period allows very accurate determinations
of the radii of star and planet (relative to the semi-major axis) and the inclination
of the orbital plane of the planet with respect to the line of sight towards the 
observer.  Clearly, one does not expect the sizes of host star and planet to vary on short 
time scales, however, the presence of a third body can change the orientation of the orbit plane and, hence, lead to a change in the observed inclination with respect to the 
celestial plane. The TrES-2 exoplanet is particularly interesting in this context.  It orbits
its host star once in 2.47~days, which itself is very much solar-like with parameters consistent with solar parameters; its age is considerable and, correspondingly, the star rotates quite slowly.  
Its close-in planet with a size of 1.2~R$_{\mathrm{Jup}}$ is among the most massive known transiting extrasolar planets (\cite{2007ApJ...664.1185H, 2007ApJ...664.1190S}). 

What makes the TrES-2 exoplanet orbits even more interesting, is the fact that an apparent inclination change has been reported by some of us in a previous paper (\cite{2009A&A...500L..45M}, henceforth 
called Paper I). The authors carefully measured several transits observed in 2006 and 2008.  Assuming a circular orbit with constant period $P$, the duration of an extrasolar planet transit in 
front of its host star depends only on the stellar and planetary radii, $R_{s}$ and $R_{p}$, and on the inclination $i$ of the orbit plane w.r.t. the sky plane. A linear best fit to the currently 
available inclination measurements yields an apparent inclination decrease of $5.1 \times 10^{-4}$~deg/day. The transit modeling both by \cite{2007ApJ...664.1185H} and \cite{2009A&A...500L..45M} shows 
the transit of TrES-2b in front of its host star to be ``grazing''. In fact, according to the latest modeling the planet occults only a portion of the host star and transits are expected to disappear 
in the time frame 2020 -- 2025, if the observed linear trend continues.  This ``grazing'' viewing geometry is particularly suitable for the detection of orbital changes, since relatively small changes 
in apparent inclination translate into relatively large changes in eclipse duration.  At the same time, a search for possible TTVs
by \cite{2009arXiv0909.1564R} has been negative; while \cite{2009arXiv0909.1564R} derive a period wobble of 57~sec for TrES-2, the statistical quality of 
their data is such that no unique periods for TTVs can be identified.

The purpose of this paper is to present new transit observations of the TrES-2 exoplanet system obtained in 2009, which are described in the first part (Sec.\ref{observationsanddatareduction}) and 
analysed (Sec. \ref{modelanalysisandresults}). In the second part (Sec.\ref{theoreticalimplications}) of our paper we present a quantitatively analysis of what kind  of gravitational effects can be 
responsible for the observed orbit changes of TrES-2b  and are consistent with all observational data of the TrES-2 system.

% of this paper is dedicated to new observations of transits in the TrES-2b exoplanet system obtained in 2009. We compare the current state of TrES-2b with that observed in 
%2008 and earlier. Additionally we took the first simultaneous four channel photometry data of the TrES-2b exoplanet using the 2.2m telescope at Calar Alto with BUSCA (\cite{1999SPIE.3649..109R}) in order 
%to study limb darkening effects in the transit curves. Since the stellar and planetary radii are not expected to change on short time scales, any observed change in transit duration is then interpreted 
%as a change in orbit inclination, and a change in orbit inclination would require, for example, the presence of a third body, such as an additional planet in the system, a non-spherical gravitational 
%field of the host star or some other effect. 

\section{Observations and data reduction}
\label{observationsanddatareduction}

We observed two transits of TrES-2 using the ephemeris suggested by \cite{2006ApJ...651L..61O} and by \cite{2007ApJ...664.1185H} from
$T_{c}(E) = 2,453,957.6358 [HJD] + E\cdot(2.47063 \ \mathrm{days})$,
\noindent
using the 1.2m Oskar L\"uhning telescope (OLT) at Hamburg Observatory and Calar Alto Observatory 2.2m telescope with BUSCA. \par
The OLT data were taken on  11 April 2009 using a $3Kx3K$ CCD with a $10'x10'$ FOV and an I-band filter as in our previous observing run (Paper I). The readout noise and the gain were 
16.37~$e^{-}$ and 1.33~$e^{-}/ADU$ respectively. With the OLT we used 60-second exposures which provided an effective time resolution of 1.17~minutes. During the observation, the airmass value ranged 
from 1.7877 to 1.0712 and the seeing was typically 2.94".\par
The Calar Alto data were taken on 28 May 2009 using BUSCA and the 2.2m telescope. BUSCA is designed to perform simultaneous observations in four individual bands with a FOV of $11' x 11'$. 
Therefore it has four individual $4Kx4K$ CCD systems which cover the ultra-violet, the blue-green, the yellow-red and the near-infrared part of the spectrum (channel a-d respectively). For 
this run we used the Str\"{o}mgren filters v (chn. a), b (chn. b),  and y (chn. c) , and  a Cousins-I filter for the near-infrared (chn. d). The readout-noise for these four detectors are 
9.09$e^{-}$, 3.50$e^{-}$, 3.72$e^{-}$, and 3.86$e^{-}$ respectively for the a,b,c, and d channels. The gain values for the same channels are 1.347$e^{-}/ADU$,  1.761$e^{-}/ADU$, 1.733$e^{-}/ADU$, and  
1.731$e^{-}/ADU$ respectively. The airmass value ranged from 1.8827 to 1.0453 and the seeing was 3.09". For the BUSCA observations we 
took 30 seconds exposure yielding an effective time resolution of 1.63~minutes. In Table \ref{tab1} we summarize the relevant observation details.

  \begin{table}

 \caption[]{Observation summary}

         \label{tab1}

         $\begin{array}{lllll}

            \hline

            \noalign{\smallskip}

             Date     &  Instrument & Filter & Airmass & Seeing \\

            \noalign{\smallskip}

            \hline

            \noalign{\smallskip}

              11/04/2009 & OLT   & I          &  1.7877 - 1.0712  &   2.94"   \\
              28/05/2009 & BUSCA & v, b, y, I &  1.8827 - 1.0453  &   3.09" \\

            \noalign{\smallskip}

            \hline

         \end{array}$

   \end{table}

For the data reduction, we used \textit{Starlink} and \textit{DaoPhot} software routines, and the \textit{MATCH} code. We perform the normal reduction tasks, bias subtraction, dark correction, and 
flat fielding on the individual data sets before applying aperture photometry on on all images. For TrES-2, we selected the aperture photometry mode using apertures centered on the target star, 
check stars, and sky background. Typical sky brightness values for the 11 April and for the 28 May were 89 and 98 ADUs, respectively, i.e., values at a level $0.008\%$ and $0.006\%$ of the star's flux, 
respectively (for I-filter). For the relative photometry we used the star U1350-10220179 as a reference star to test and calibrate the light curve. For the data analysis presented in this paper we did 
not use additional check stars, but note that we already checked this star for constancy in the Paper I. To estimate the magnitude errors, we followed 
\cite{2001phot.work....1H} and the same procedure as described in our first paper (Paper I) to obtain better relative results. Our final relative photometry is presented in Table \ref{tab2}, which is 
available in its entirety in machine-readable form in the electronic version of this paper.

  \begin{table}
  
      \caption[]{Relative Photometry data.}

         \label{tab2}

         $\begin{array}{llll}

            \hline

            \noalign{\smallskip}

             HJD     &  Relative$ $Flux & Uncertainty & Flag\\

            \noalign{\smallskip}

            \hline

            \noalign{\smallskip}

              2454933.44031 & 0.99172 & 0.0037 & OLT-I  \\
              2454933.44091 & 0.99204 & 0.0038 & OLT-I  \\
              2454933.44221 & 1.00514 & 0.0039 & OLT-I \\
              2454933.44281 & 1.00314 & 0.0038 & OLT-I \\
              2454933.44341 & 0.99458 & 0.0035 & OLT-I   \\
              2454933.44401 & 0.99745 & 0.0037 & OLT-I  \\

            \noalign{\smallskip}

            \hline

         \end{array}$

Note : Table 2: Relative photometry vs. time; note that the complete table is available only electronically. The time stamps refer to the Heliocentric Julian Date (HJD) at the middle of each exposure. The 'Flag' column refers to the telescope and filter.

   \end{table}

\section{Model analysis and results}
\label{modelanalysisandresults}

In our model analysis we proceeded in exactly the same fashion as described in Paper I.  Note that the assumption of circularity appears to apply very well to
TrES-2b (\cite{2009IAUS..253..536O}; \cite{2006ApJ...651L..61O}); the assumption of constant period and hence constant semi-major axis will be adressed in
section 4.  For our modelling
%period of the planet remain constant, which again is supported by the available observational data (see section 4; \cite{2009AN....330..459R})}. 
we specifically assumed the values  $R_{s}=1.003R_{o}$, 
$R_{p}=1.222\ R_{J}$, $P=2.470621$ days, $\alpha = 0.0367$ AU for the radii of star and planet, their period and the orbit radius respectively.  All limb darkening coefficients were taken from 
\cite{2004A&A...428.1001C}, and for the OLT data we used the same values as in Paper I, viz., $u_{1}$ = 0.318 and $u_{2}$ = 0.295 for the I filter, as denoted by S1 in Table \ref{tab3}.

\subsection{OLT data \& modeling}

    \begin{figure}[ht]
    \centering
    \includegraphics[width=6.0cm]{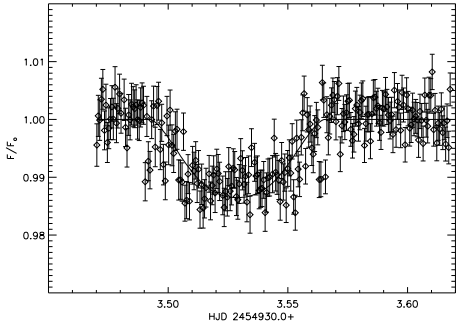}
   \caption{Observed TrES-2 light curves and model fits for the light curve taken with the 1.2m OLT at Hamburg Observatory taken in April 2009.}
\label{fig1}

\end{figure}

    \begin{figure}[ht]

    \centering

    \includegraphics[width=7.2cm]{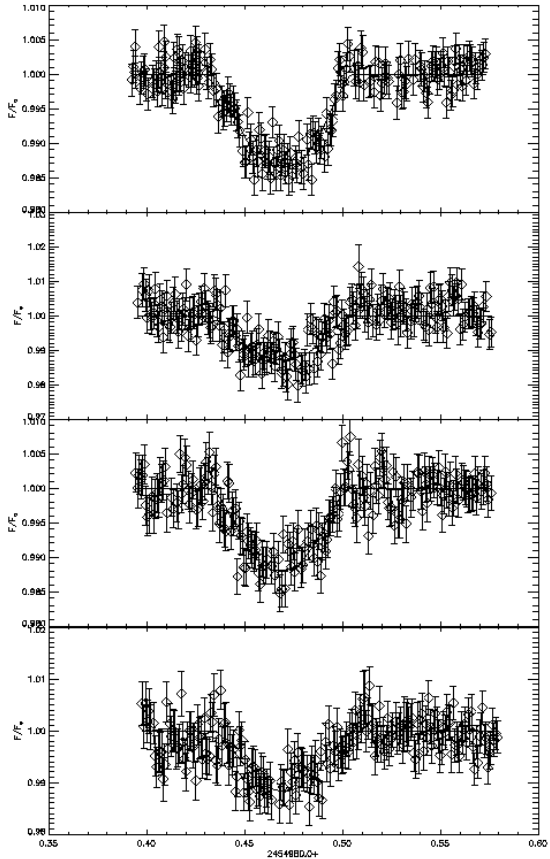}

   \caption{Four light curves and corresponding model fits taken in May 2009 with the 2.2m telescope at Calar Alto using BUSCA.}

\label{fig2}

\end{figure}

\begin{figure}[ht]

    \centering
    \includegraphics[width=6.0cm]{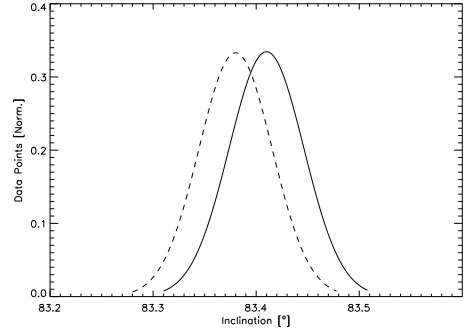}
   \caption{Inclination distribution for the OLT data using 1000 Monte Carlo simulations for the September 2008 (solid curve) and April 2009 (dashed curve); see the text for details.}
\label{fig3}
\end{figure}

Our new OLT data from 11 April 2009 were taken with the same instrumental setup as our previous data taken on 18 September 2008. The final transit light curve with the best fit model is shown in 
Fig.\ref{fig1}, the reduced light curve data are provided in Table \ref{tab1}. Keeping the planetary and stellar radii and the limb darkening coefficients fixed (at the above values), we determine 
the best fit inclination and the central transit time $T_{c}$ with our transit model using the $\chi^{2}$ method; the thus obtained fit results are listed in Table \ref{tab3}. The errors in the derived 
fit parameters are assessed with a bootstrap method explained in detail in Paper I, however, we do not use random residuals for the new model, but we circularly shift the residuals 
after the model substraction to produce new light curves for the bootstrapped data following \cite{2008A&A...487L...5A}. In Tab. \ref{tab3} we also provide corrected central times and errors for those
transists already reported in paper I, since due to some typos and mistakes the numbers quoted for central time and their error are unreliable (five first lines of Table \ref{tab3}).
In Fig. \ref{fig3} we plot the thus obtained inclination angle distribution for epoch April 2009 (solid curve) compared to 
that obtained in September 2008 (dashed curve). While the best fit inclination of $i=83.38^{\circ}$ differs from that measured in September 2008 ($i=83.42^{\circ}$), the errors are clearly so large that 
we cannot claim a reduction in inclination from our OLT data (taken with the same instrumental setup, i.e, in September 2008 and April 2009) alone, however, is clearly consistent with such 
a reduction.

   \begin{table}

      \caption[]{Individual values of duration, inclination, $\chi^{2}$ values and limb darkening coefficients from five light curve fits; units for duration and errors are minutes and for inclination and 
errors are degrees. The OLT light curve has 268 points, the I-filter light curve has 198, the y-filter light curve has 208, the b-filter light curve 198 and the v-filter light curve 162 points 
respectively.}

         \label{tab3}

         $\begin{array}{llllllll}

            \hline

            \noalign{\smallskip}

            \textit{Tc}$ $time$ $[HJD] &  Duration & Errors & Inclin. & Errors & \chi^{2} value & LDL \\

            \noalign{\smallskip}

             \hline

            3989.7529\pm0.00069  & 110.308 & 0.432 &  83.59  &  0.019 &  432.1   \\
            3994.6939\pm0.00066  & 109.230 & 0.448 &  83.56  &  0.019 &  296.8 \\
            4041.6358\pm0.00070  & 109.025 & 0.430 &  83.55  &  0.019 &  290.6 \\ 
            4607.4036\pm0.00072  & 106.620 & 0.883 &  83.44  &  0.036 &  179.1 \\
            4728.4640\pm0.00071  & 106.112 & 0.870 &  83.43  &  0.036 &  190.1 \\

            \hline

            \noalign{\smallskip}

            4933.5274\pm0.00076 & 105.077 & 0.964 &  83.38  &  0.039 &  296.1 & S1 \\
            4980.4675\pm0.00068 & 105.056 & 0.848 &  83.38  &  0.034 &  181.7 & S1\\
            4980.4679\pm0.00090 & 104.748 & 1.076 &  83.37  &  0.043 &  202.2 & S2\\
            4980.4667\pm0.00082 & 103.832 & 1.021 &  83.33  &  0.041 &  186.0 & S3\\
            4980.4678\pm0.00100 & 104.363 & 1.194 &  83.35  &  0.048 &  195.0 & S4\\

            \noalign{\smallskip}

            \hline

            4980.4675\pm0.00060 & 104.522 &       &  83.36   &  0.030     \\

            \hline\hline

         \end{array}$

   \end{table}

\subsection{BUSCA data and modeling}

Our BUSCA data have the great advantage of providing simultaneously measured multicolor data, which allows us to demonstrate that limb darkening is correctly modelled and does not affect the fitted 
inclinations and stellar radii. The limb darkening coefficients used for the analysis of the BUSCA data were also taken from \cite{2004A&A...428.1001C}; we specifically used $u_{1}$ = 0.318 and 
$u_{2}$ = 0.295 for the I filter (S1), i.e., the same values as for the OLT, $u_{1}$ = 0.4501 and $u_{2}$ = 0.286 for the y filter (S2), $u_{1}$ = 0.5758 and $u_{2}$ = 0.235 for the b filter (S3), 
$u_{1}$ = 0.7371 and $u_{2}$ = 0.1117 for the v filter (S4) respectively. The data reduction and analysis was performed in exactly the same fashion as for the OLT data, we also used the same comparison star 
U1350-10220179; the reduced light curve data are also provided in Table \ref{tab2}. 
The modelling of multicolor data for extrasolar planet transits needs some explanation.  In our modelling the host star's radius is assumed to be wavelength independent. Since stars do not have solid 
surfaces, the question arises how much the stellar radius $R_{*}$ does actually change with the wavelength. This issue has been extensively studied in the solar context, where the limb of the Sun can be 
directly observed as a function of wavelength \citep{1995SoPh..156....7N}.  Basically the photospheric height at an optical depth of unity is determined by the ratio between pressure and the absorption 
coefficient, and for the Sun, \cite{1995SoPh..156....7N} derives a maximal radius change of 0.12 arcsec between 3000 and 10000 \AA , which corresponds to about 100 km. If we assume similar photospheric 
parameters in the TrES-2 host star, which appears reasonable since TrES-2 is a G0V star, we deduce a 
wavelength-dependent radius change on the order of 100 km, which is far below a percent of the 
planetary radius and thus not detectable. Therefore in our multi-band data modelling we can safely fix the radius of the star and the radius of the planet in a wavelength-independent fashion.\\
We thus kept the stellar and planetary fixed at the above values, treated all BUSCA channels as independent observations and fitted the light curves -- as usual -- by adjusting inclination and central
transit time. The filter light curves are shown together with the so obtained model fits in Fig. \ref{fig2}, the model fit parameters are again summarized in Table \ref{tab3}.  We emphasize that we obtain 
good and consistent fits for all light curves with the chosen set of limb darkening coefficients, thus demonstrating our capability to correctly model multicolor light curves.\\

   \begin{table}

      \caption[]{$\chi^{2}$ tests for two different inclination values and the $\chi^{2}$ errors after 1000 Monte Carlo simulations.}

         \label{tab4}

         $\begin{array}{llllll}

            \hline

            \noalign{\smallskip}

            \textit Inclination$ $[^{\circ}] &  OLT & B-I & B-y & B-b & B-v  \\

            \noalign{\smallskip}

             \hline
            \noalign{\smallskip}

            83.57$ $(Holman)       & 304.9& 199.4& 207.8&  216.0& 206.3 \\
            \sigma_{\chi^{2}}$ $(Holman)      &16.12 &14.04& 16.16&21.05& 14.26 \\
            83.36$ $(this$ $paper) & 213.8& 173.44 & 164.3  &  180.0 &  131.1 \\
            \sigma_{\chi^{2}}$ $(this$ $paper)&11.96 &11.11 & 9.96& 12.01 & 8.77 \\

%            83.57$ $(Holman)       & 304.9 & 199.4 & 207.8  &  216.0 &  206.2 \\
%            83.36$ $(this$ $paper) & 296.2 & 182.3 & 202.3  &  187.0 &  195.2 \\
%            \Delta \chi^{2}        & 8.7 & 17.1 & 5.5  &  29.0 &  10.8 \\

            \noalign{\smallskip}

            \hline
         \end{array}$
   \end{table}

Since the BUSCA data are recorded simultaneously, it is clear that the light curves in the four BUSCA channels must actually be described by the same values of inclination and transit duration. 
We therefore simultaneously fit all the four light curves, leaving free as fit parameters only the central time $T_{c}$ and inclination $i$. With this approach we find an average inclination of 
$i=83.36^{\circ} \pm 0.03^{\circ}$, which is consistent with our spring OLT data and also suggests that the inclination in spring 2009 has further decreased as compared to our 2008 data.

\begin{figure}[ht]

\centering

\includegraphics[width=8cm,angle=0,clip=true]{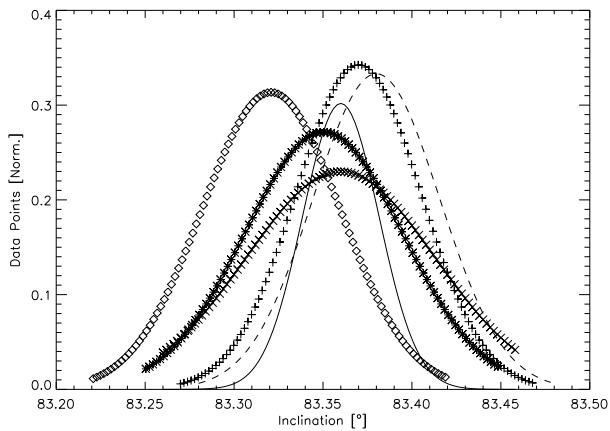}

\caption{Inclination distribution derived 1000 Monte Carlo simulations for the multi-band BUSCA observations data in the I (crosses), y (stars), b (diamonds) and v (x-symbol) filter bands; the mean 
inclination distribution from the four BUSCA lightcurves is shown as solid line, the inclination distribution derived from the OLT data taken in April 2009 is also shown for comparison (dashed line).} 

\label{fig4}

\end{figure}

\subsection{Joint modeling}

Using all our data we can check whether our assumption of the wavelength-independence of the radius is consistent with the observations. For this consistency test we kept the inclination value fixed at 
$i=83.36^{\circ}$ and fitted only the radius of the star $R_{s}$ and $T_{c}$. The errors on $R_{s}$ were again assessed by a Monte Carlo simulation as described in Paper I and the distribution of the thus 
derived stellar radius values $R_{s}$ is shown in Fig. \ref{fig4}; as apparent from  Fig. \ref{fig4}, all BUSCA channels are consistent with the same stellar radius as -- of course -- expected from theory since any pulsations of a main-sequence star are not expected to lead to any observable radius changes.

The crucial issue about the TrES-2 exoplanet is of course the constancy or variability of its orbit inclination.  Our new BUSCA and OLT data clearly support a further decrease in orbit inclination and
hence decrease in transit duration.  In order to demonstrate the magnitude of the effect, we performed one more sequence of fits, this time keeping all  physical parameters fixed and fitting only 
the central transit times $T_{c}$ using 1000 Monte Carlo realisations and studying the resulting distribution in $\chi^2$; for the inclination we assumed for, first, the value $i=83.57^{\circ}$ 
(as derived by Holman et al. for their 2006 data) and the value $i=83.36^{\circ}$ (this paper for the 
2009 data). The fit results (in terms of obtained $ \chi ^2$ values) are summarised ed in Tab. \ref{tab4}, which shows that for all (independent) data sets the lower inclination values yield 
smaller  $ \chi ^2$-values; for some filter pairs the thus obtained improvement is extremely significant.

\begin{figure}[ht]

\centering

\includegraphics[width=8cm,angle=0,clip=true]{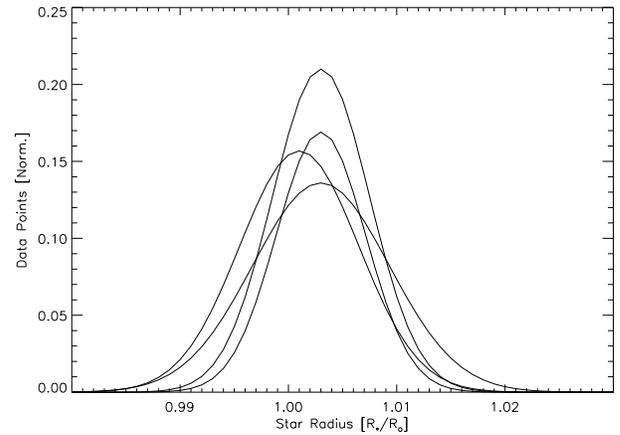}

\caption{Stellar Radius distribution $(R_{*})$, derived from 1000 Monte Carlo simulations in four different filters (from higher to lower peak, I, y, b and v filter respectively). The overlap of the 
curves shows that all colors can be explained with the same stellar radius as suggested by theory (see \cite{1995SoPh..156....7N}).}

\label{fig5}

\end{figure}

\subsection{Inclination changes}

\begin{figure}[ht]
   \centering
   \includegraphics[width=8cm,angle=0,clip=true]{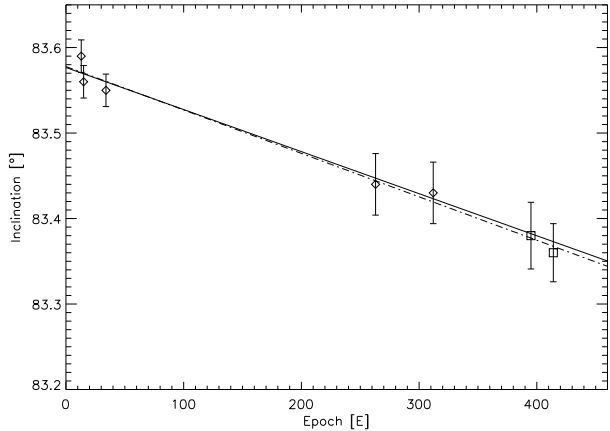}
\caption{Epoch versus inclination together with a linear fit to the currently available data; the diamond points are those taken in 2006 by \cite{2007ApJ...664.1185H}, and those taken in 2008 and reported 
in Paper I. The square points are derived from our new observations taken in April and May 2009. The solid lines showing two linear fits, from the first paper (dashed line) and the fit from 
the present paper (solid line).}
    \label{fig10}
    \end{figure}

In Fig. \ref{fig10} we plot all our current inclination vs. epoch data 
together with a linear fit to all data; a formal regression analysis yields for the time evolution of the inclination ($i=i_{o}+a \cdot (Epoch)$, $i_{o}=83.5778$, $a=0.00051$). In Paper I we noted the 
inclination decrease and predicted inclination values below the first transit threshold ($i_{min,1}<83.417^{\circ}$) after October 2008.  Both the new OLT data set and all BUSCA channel observations  yield 
inclinations below the first transit threshold. While the error in a given transit light curve is typically on the order of 
0.04$^{\circ}$ for $i$, we consider it quite unlikely that 5 independent 
measurements all yield only downward excursions. We therefore consider the decrease in inclination between fall 2008 and spring 2009 as significant, conclude that the inclination in the TrES-2 system 
is very likely below the first transit threshold, and predict the inclinations to decrease further; also, the transit depths should become more and more shallow since the exoplanet eclipses less and 
less stellar surface.

\subsection{Period changes}

\begin{figure}[ht]
   \centering
   \includegraphics[width=8cm,angle=0,clip=true]{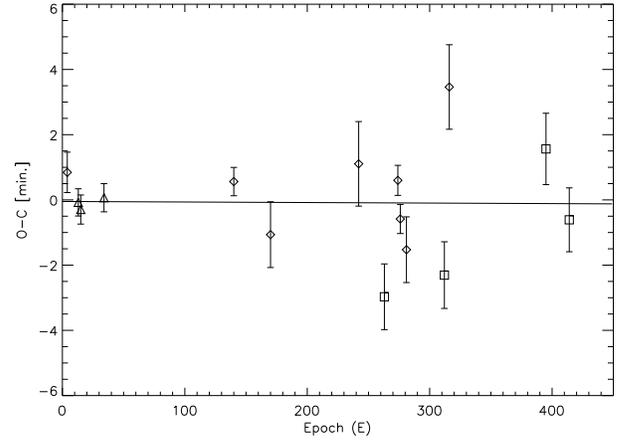}
\caption{O-C values versus epoch including the transits observed by  Holman et al., Rabus et al. and our data denoted by  triangles, squares and diamonds respectively.}
%\textbf{Down:} $\chi^{2}$ vs. $\dot{P}$. The best fit solution $\dot{P} \sim 0$ does not leave us any assumptions for period changes.}
    \label{fig6}
    \end{figure}

The observed change in orbit inclination is in marked contrast to the period of TrES-2b.  While possible TTVs in TrES-2b have
been studied by \cite{2009arXiv0909.1564R} we investigate the long-term stability of the period of TrES-2b. 
From our seven transit measurements (plus five more data points of \cite{2009arXiv0909.1564R}) spanning about 
$\sim$ 400 eclipses we created a new  O-C diagram (cf., Fig. \ref{fig6}); note that we refrained from using the transit times dicussed
by  \cite{2009AN....330..459R}, since these transits were taken with rather inhomogeneous instruments and sometimes only partial transit coverage. 
%More detailed analysis has been done by \cite{2009arXiv0909.1564R} who show that there is no evidence 
%for timing variation in TrES-2 exoplanet. 
For our fit we used a modified epoch equation $HJD_{c}=HJD_{o}+E \cdot P$, where we set 
$P=P_{o}+\dot{P}(t-HJD_{o})$ and explicitly allow a non-constant period $P$. We apply a $\chi ^{2}$ fit to find the 
best fit values for $\dot{P}$, $P_{o}$, and $HJD_{o}$.  With this approach we find a best fit period change of  $\dot{P}$ = 5 $\times$ 10$^{-9}$, however, 
carrying out the same analysis keeping a fixed period shows that
the fit improvement due to the introduction of a non-zero $\dot{P}$ is insignificant.  We then find as best fit values  $P_{o}=2.47061$ and $HJD_{o} =2453957.6350$ conforming to the values
derived by \cite{2009arXiv0909.1564R}.  Thus, the period of TrES-2b is constant, and any possible period change over the last three years must be less than about 1~sec.

\section{Theoretical implications of observed inclination change}
\label{theoreticalimplications}

The results of our data presented in the preceding sections strenghten our confidence that the observed inclination changes do
in fact correspond to a real physical phenomenon.  Assuming now the reality of the observed inclination change of $\Delta i \sim 0.075^{\circ}/\textrm{yr}$, given the 
constancy of the period and the absence of TTVs at a level of $\approx$100~sec we discuss in the following a physical scenario
consistent with these observational findings.  We specifically argue that the apparent inclination change should be interpreted
as a nodal regression and then proceed to examine an oblate host star and the existence of an additional perturbing object in the system as possible causes for the change of the orbital parameters of Tres-2b.

\subsection{Inclination change or nodal regression?}

It is important to realize that the reported apparent change of inclination refers to the orientation of the TrES-2 orbit plane with respect to the observer's tangential plane.  It is well known that the 
z-component of angular momentum for orbits in an azimuthally symmetric potential is constant, resulting in a constant value of inclination.  An oblate star (cf., Sec.~\ref{oblate}) or the averaged 
potential of a third body (cf., Sec.~\ref{thirdbody}) naturally lead to such potentials, thus yielding orbits precessing at (more or less) constant inclination as also realized by 
\cite{2002ApJ...564...60M}.  Such a precession would cause an apparent inclination change, however, physically this would have to be interpreted as nodal regression at fixed orbit inclination. 
To interpret the observations, one has to relate the rate of nodal regression to the rate of apparent inclination change. 

\begin{figure}[ht]
\centering
\includegraphics[width=6.0cm]{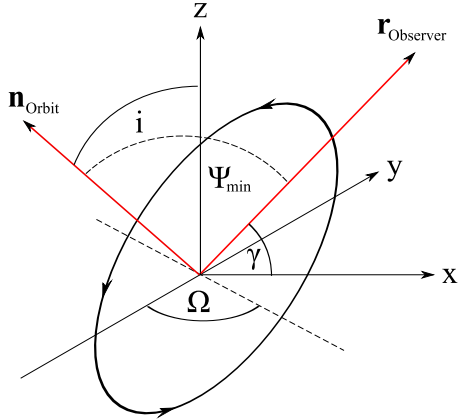}
\caption{System geometry (see text for details): The orbit plane with normal vector $\vec{n}_{\mathrm{Orbit}}$ is inclined relative to the x-y plane by an inclination $i$, the observer (towards $\vec{r}_{\mathrm{Observer}}$) is within the x-z plane.}
\label{fig7}
\end{figure}

Consider a (massless) planet of radius $R_p$ orbiting a star of radius $R_{*}$ at some distance $d$.  Let the planet's orbit lie in a plane with a fixed inclination $i$ relative to the x-y plane, which we take as invariant plane.  Let an observer be located in the
x-z plane with some elevation $\gamma$, reckoned from the positive x-axis.  The
line of the ascending node in the x-y plane is denoted by the angle $\Omega$, with 
$\Omega = 0$ implying the ascending node pointed along the negative y-axis (see Fig.~\ref{fig7}). 
Let in the thus defined geometry the angle $\Psi$ denote the angle between planet and observer as seen from the central star. For each system configuration defined by the angles ($\gamma, i, \Omega$) there is a minimal angle $\Psi_{\mathrm{min}}$ between orbit normal and observer obtained in each planetary orbit, which can be computed from
\begin{equation}
%\cos{\Psi_{\mathrm{min}}} = 
%\sqrt{(\cos{\gamma}\cos{i}\sin{\alpha}+\sin{\gamma}\sin{i})^2+(\cos{\gamma}\cos{\alpha})^2}.
\cos{\Psi_{\mathrm{min}}} = \vec{n}_{\mathrm{Orbit}} \cdot \vec{r}_{\mathrm{Observer}} =
-\cos{\gamma}\sin{i}\cos{\Omega}+\sin{\gamma}\cos{i}.
\label{iobs}
\end{equation} 
A transit takes place when
\begin{equation}
|\cos{\Psi_{\mathrm{min}}}| \le (R_p+R_*)/d ,
\end{equation} 
and from the geometry it is clear that the observed inclination $i_{\mathrm{obs}}$, i.e.,
the parameter that can be derived from a transit light curve is identical to 
$\Psi_{\mathrm{min}}$.  Setting then $\Psi_{\mathrm{min}} = i_{\mathrm{obs}}$ and
differentiating Eq.~\ref{iobs} with respect to time we obtain
\begin{equation}
%\frac{\mathrm{d} i_{\mathrm{obs}}}{\mathrm{d}t} = - \frac{\cos{\alpha}\cos{\gamma}\sin{i}
%\left( \cos{\gamma}\sin{i}\sin{\alpha} -\sin{\gamma}\cos{i} \right) }{\sin{i_{\mathrm{obs}}}\cos{i_{\mathrm{obs}}}}\frac{\mathrm{d} \alpha} {\mathrm{d}t}.
\frac {\mathrm{d} i_{\mathrm{obs}}} {\mathrm{d}t} = 
- \frac {\sin{\Omega} \cos{\gamma} \sin{i}} { \sin{i_{\mathrm{obs}}}} \frac {\mathrm{d} \Omega} {\mathrm{d}t}.
\label{adot}
\end{equation}
 
\begin{figure}[ht]
\centering
\includegraphics[width=8.0cm]{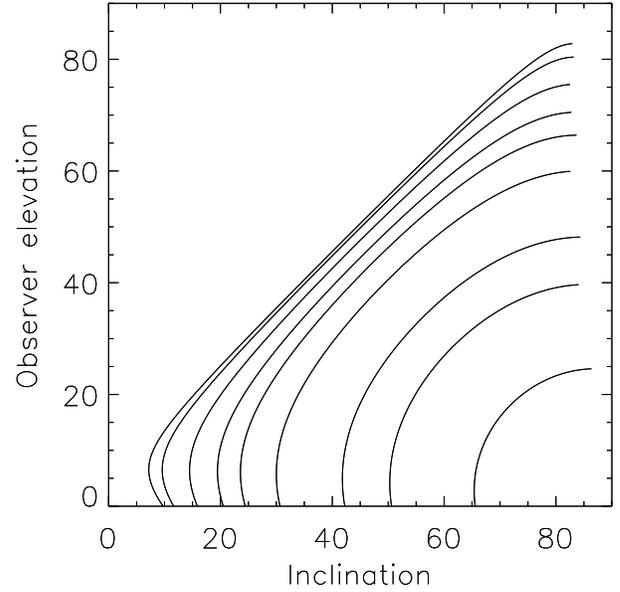}
\caption{Linear coefficient of Eq.~\ref{adot} between nodal regression and observed inclination change as a function of view geometry 
(cf., Eq.~\ref{adot}), computed for $i_{\mathrm{obs}}$ = 83.38$^{\circ}$ as applicable for TrES-2; the plotted contour levels 
denote values of 1.1, 1.3, 1.5, 2., 2.5, 5.,4.,6.,8. from right bottom up.}
\label{fig8}
\end{figure}

Eq.~\ref{adot} relates the nodal regression of the orbit to its corresponding observed apparent rate of inclination change $\mathrm{d}i_{\mathrm{obs}}/\mathrm{d}t$, given the fixed inclination 
$i$ relative to the x-y plane. 
% Eq.\ref{adot} relates an observed inclination rate $\mathrm{d}i_{\mathrm{obs}}/\mathrm{d}t$ to the corresponding nodal regression rate.  
Since transit observations yield
very precise values of  $i_{\mathrm{obs}}$, the required ascending node $\Omega$ and its rate of change can be computed, once the orbit geometry through the angles $\gamma$ and $i_{\mathrm{obs}}$ is specified.
In Fig.~\ref{fig8} we show a contour plot of the linear coefficient of Eq.~\ref{adot} between nodal regression and observed inclination change as a function of orbit geometry. Note that the 
apparent change of inclination due to the nodal regression does vanish for $i=0$. Physically it is clear that a perturbing planet in a coplanar orbit cannot exercise a torque and therefore cannot
cause the observed inclination variation. We will therefore always assume $i\neq0$ in the following.
    
\subsection{Oblate host star}
\label{oblate}

%\subsection{Theory}

We first consider the possibility that the TrES-2 host star is oblate.
The motion of a planet around an oblate host star is equivalent to that of an artifical satellite orbiting the Earth, a problem intensely studied over the last decades 
and well understood \cite{2005Obs...125..341C}.  The potential 
$U(r,\phi)$ of an axisymmetric body of mass $M$ and radius $R$ can be expressed as a power series involving the so-called harmonic coefficients.  In second order one approximates
the potential  $U(r,\phi)$  as
\begin{equation}
U(r, \phi ) = \frac{G M} {r} \left[1 - J_2 \left(\frac{R} {r}\right)^2 \frac{1} {2} \left(3\sin^2{\phi} - 1\right)\right],
\end{equation}
%\begin{equation}
%U(r,\phi ) = \frac{G M} {r} \left[1 - \sum_{n=2}^{\infty} J_n \left(\frac{R} {r}\right)^n P_n(\sin{\phi})\right],
%\end{equation}
where r is the radial distance from the body's center, $\phi$ is latitude above the equator and $G$ denotes the gravitational constant.
%For the quite  rapidly rotating Earth one finds \mbox{$J_2 = 0.00108$}, all other harmonic coefficients are smaller by at least three orders of magnitude, thus justifying the 
%neglect of all higher orders, and we assume similar conditions for the TrES-2 host star.
Clearly, the perturbing term in the potential is proportional to $J_2$ and a perturbation calculation yields
as first order secular perturbation the angular velocity of the ascending node as
%\begin{equation}
%\frac {d \Omega} {dt} = - \frac {3}{2} \frac {J_2 R^2} {a^2 (1-e^2)^2} n cos(i),
%\end{equation}
\begin{equation}
\label{eq:firstorderJ2}
\frac {\mathrm{d} \Omega} {\mathrm{d}t} = - \frac {3}{2} \frac {J_2 R^2} {(1-e^2)^2} \cos{i} \left(\frac {2 \pi} {P}\right)^{7/3} \frac {1} {(G M_{\mathrm{total}})^{2/3}},
\end{equation}
where $e$ and $i$ denote the eccentricity and inclination of the orbiting body, $P$ its period,
$M_{\mathrm{total}}$ the sum of the masses of planet and host star and the validity of Kepler's third law has 
been assumed.

%\subsection{Application to TrES-2}

Interpreting the observed inclination change as nodal regression due to an oblate host star, we can compute a lower bound on the 
required harmonic coefficient $J_2$ assuming $e = 0$. Therefore, we combine Eqs.~\ref{adot} and~\ref{eq:firstorderJ2} to obtain an expression for $J_2$. Excluding pathological cases like $i,\Omega=0$ and 
$\gamma=\pi/2$ and neglecting the planetary mass in Eq.~\ref{eq:firstorderJ2} we find for any given set of parameters $(\Omega,\gamma,i)$ 
%Therefore, we assume an infinitesimally small inclination ($\cos{i} \approx 1$). Neglecting also the planetary mass in Eq.\ref{eq:firstorderJ2} yields}
%\begin{equation}
%J_2^{\mathrm{min}} = \frac {2}{3} \frac {\mathrm{d} \Omega} {\mathrm{d}t} \left(\frac{P}{2\pi}\right)^{7/3} \frac {(G M_{\mathrm{host}})^{2/3}} {R^2},
%\end{equation}
\begin{equation}
J_2 \geq J_2^{\mathrm{min}} = \frac{2}{3} \frac{\mathrm{d} i_{\mathrm{obs}}} {\mathrm{d}t}  \left(\frac{P}{2\pi}\right)^{7/3} \frac {(G M_{\mathrm{host}})^{2/3}} {R^2} \sin{i_{\mathrm{obs}}},
\end{equation}
where $M_{\mathrm{host}}$ denotes the mass of the host star.
Mass, radius, inclination and period of the TrES-2 system are well known, and using the measured 
nodal regression we find $J_{2,\mathrm{TrES-2}} \approx 1.4 \times 10^{-4}$, i.e., a value
smaller than that of the Earth ($J_{2,\oplus} = 0.00108$) by an order of magnitude, but considerably larger than
that of the Sun, which is usually taken as $J_{2,\odot} \approx 3 - 6 \times 10^{-7}$ \cite{2001SoPh..198..223R}.  Since the host star
of TrES-2 is a slow rotator very similar to the Sun \cite{2007ApJ...664.1190S}, we expect 
similarly small $J_2$ values in contrast to our requirements.
We therefore conclude that oblateness of the host star cannot be the cause for the observed orbit variations.

\subsection{Perturbation by a third body}
\label{thirdbody}
%\subsubsection{Theory}

An alternative possibility to explain the observed orbit variations of the TrES-2 exoplanet would be the interaction
with other planets in the system.  Let us therefore assume the existence of such an additional perturbing planet of
mass $m_p$, circling its host star of mass $m_0$ with period $P_p$ at distance $r_p$ 
located further out compared to the known transiting TrES-2 exoplanet.  
This three-body problem has been considered in past in the context of triple systems (\cite{1991ApJ...375..314K}, \cite{2006AJ....131..994L}) and
the problem of artificial Earth satellites, whose orbits are perturbed by
the Moon.
In lowest order, the perturbing gravitational potential $R_2$ onto the inner planet with mass $m$ and distance $r$ is given by the expression
%\begin{equation}
%R =  \frac{G M_p} {r_p} \sum_{n=2}^{\infty} R_n = \frac{G M_p} {r_p} \sum_{n=2}^{\infty} \left(\frac{r} {r_p}\right)^n P_n(\cos{S}),
%\end{equation}
{\begin{equation}
R_2 = \frac{m_p} {m_p + m_o} \left(\frac{2 \pi a} {P_p}\right)^2 \left(\frac{a_p} {r_p}\right)^3 \left(\frac{r} {a}\right)^2 \frac{3 \cos^2{S} - 1} {2},
\end{equation}
where the angle $S$ denotes the elongation between the perturbed and perturbing planet as seen from the host star,
$a$ and $a_p$ denote their respective semi-major axes and the validity of Kepler's third law has been assumed. 
Note that in this approach
the perturbed body is assumed to be massless, implying that its perturbations onto the perturbing body are ignored.
Next one needs to insert the orbital elements of the two bodies and, since we are interested only in secular variations, average
over both the periods of the perturbed and perturbing planet. This is the so-called double-averaging method (\cite{2003JGCD...26..27B}), which, however, in more or less the same form has also
been applied by \cite{2006AJ....131..994L} and \cite{1967nmds.conf..221K}.
Denoting by $e$ the eccentricity of the perturbed planet, by $\omega$ the longitude of the periastron and by $i$ the angle between the two orbital planes,
one obtains after some lengthy computation (see \cite{1967nmds.conf..221K})

\begin{equation}
R_2 = \frac{m_p} {m_p + m_o} \left(\frac{\pi a} {2 P_p} \right)^2 
\times K_0(i,e,\omega),
\end{equation}
with the auxiliary function $K_0(i,e,\omega)$ given by
$$
K_0(e,i,\omega) = \left(6 \cos^2{i} - 2 \right) + e^2 \left( 9 \cos^2{i} -3 \right) + 15 e^2 \sin^2{i} \cos{2 \omega}.
$$
The partial derivatives of $R_2$ with respect to the orbital elements are needed in the so-called Lagrangian planetary equations to derive
the variations of the orbital elements of the perturbed body.  One specifically finds for the motion of the ascending node
\begin{equation}
\frac{\mathrm{d} \Omega} {\mathrm{d}t} = \frac{m_p} {m_p + m_o} \frac{3 \pi} {4} \frac{P} {P_p^2} \frac{\cos{i}} {\sqrt{1-e^2}} \times K_1(e,\omega),
\label{omegadot}
\end{equation}
where the auxiliary function $K_1(e,\omega)$ is defined through
\begin{equation}
K_1(e,\omega) = 5 e^2 \cos{2 \omega} - 3e^2 -2
\end{equation}
and for the rate of change of inclination
\begin{equation}
\frac{\mathrm{d} i} {\mathrm{d}t} = -e^2 \frac{m_p} {m_p + m_o} \frac{15 \pi} {8} \frac{P} {P_p^2} \frac{1} {\sqrt{1-e^2}} \sin{(2i)} \sin{(2\omega)}.
\label{idot}
\end{equation}
As is obvious from Eq.~\ref{idot}, the rate of change of inclination in low eccentricity systems is very small
and we therefore set $e$ = 0.
Assuming next a near coplanar geometry, i.e., setting $\cos{i} \approx 1$, we can simplify Eq.~\ref{omegadot} as
\begin{equation}
\label{omegadotsimple}
\frac{\mathrm{d} \Omega} {\mathrm{d}t} = - \frac{m_p} {m_p + m_o} \frac{3 \pi} {2} \frac{P} {P_p^2} . 
\end{equation}
If we assume a host star mass and interpret the observed inclination change as the rate of nodal regression via Eq.~\ref{adot}, 
Eq.~\ref{omegadotsimple} relates the unknown planet mass $m_p$ to its orbital period $P_p$.

\subsubsection{Sanity check:  Application to the Solar System}

The use of Eq.~\ref{omegadotsimple} involves several simplifications. Thus, it is legitimate to ask, if we are justified in expecting Eq.~\ref{omegadotsimple} to describe reality. As a sanity 
check we apply Eq.~\ref{omegadotsimple}
to our solar system.  Consider first the motion of the Moon around the Earth, i.e., $P = 27.3$~d, which is 
perturbed by the Sun, i.e., $P_p = 365.25$~d.   Since for that case our nomenclature requires \mbox{$m_p \gg m_o$} and $i \sim 5.1^{\circ}$,
we find from Eq.~\ref{omegadotsimple} a time of 17.83 years for the nodes to complete a full circle, which agrees
well with the canonical value of 18.6 years for the lunar orbit.  In the lunar case it is clear that the Sun with its large mass
and close proximity (compared to Jupiter) is by far the largest perturber of the Earth-Moon two-body system and this situation is exactly
the situation described by theory.

Consider next the the perturbations caused by the outer planets of our solar system.  Considering, for example, Venus, we can compute
the perturbations caused by the planets Earth, Mars, Jupiter, Saturn and Uranus.  Since the perturbation strength scales
by the ratio $m_p P_p^{-2}$, we can set this value to unity for the Earth and compute values of 0.03, 2.26, 0.11 and 0.002
for Mars through Uranus respectively.  So clearly, Venus is perturbed by several planets, but the perturbations by Jupiter
are strongest.  We therefore expect that our simple approach is not appropriate.  We further note that among the outer
solar system planets long period perturbations and resonances occur, which are not described by Eq.~\ref{omegadotsimple}.  If we 
nevertheless compute the nodal regression for Venus caused by Jupiter using
Eq.~\ref{omegadotsimple}, we find a nodal regression of 0.1$^{\circ}$/cty for Venus, and 0.3$^{\circ}$/cty for Mars.
Using the orbital elements computed by \cite{1994A&A...282..663S} and calculating the nodal regressions of Venus and Mars 
in the orbit plane of Jupiter we find values smaller than the true values,
but at least, they computed values in the right order of magnitude and do not lead to an overprediction of the expected effects.

%An overlooked consequence of the orbital element evolution of the solar system planets is as follows:  Consider an
%hypothetical observer located far away from the solar system fortuitously located in the plane of the ecliptic at exactly the
%longitude of Jupiter's, say, ascending node.  By construction, such an observer could discover that the Sun is orbited not only 
%by an Earth-like planet, but orbited by a ``cold'' Jupiter; such a Jupiter transit, lasting for about 1.4~days,  would be observed every 11.86~years, and by correct light curve modeling, this observer would observe an apparent orbit inclination of 90$^{\circ}$.   Repeating this observation 12~years later, this hypothetical 
%observer would deduce an inclination change of 72~arcsec between two observed Jupiter transits since 
%Jupiter's nodal regression 6~arcsec/yr in this observer's plane.  Since the Sun has an
%apparent diameter of only 5.8~arcmin as seen from Jupiter, only four such transits would be observable during a given ``transit season''.
%The geometrical situation is in fact the same as for the Saros cycle for
%solar eclipses.  A more favorable situation for an hypothetical extrasolar Jupiter observer would of course arise by moving into the% current plane of the Jupiter 
%orbit.  In summary, the known nodal regressions in our solar system would lead to transient transit phenomena for
%hypothetical transit observers located outside the solar system.

\subsubsection{Sanity check: Application to V~907~Sco}

From archival studies \cite{1999AJ....117..541L} report the existence  of transient eclipses in the triple star V~907~Sco.  According to
\cite{1999AJ....117..541L} this system is composed out of a short-period (P$_{\mathrm{short}}$ = 3.78~days) binary containing two main sequence
stars of spectral type $\sim$ A0 and mass ratio 0.9, orbited by a lower mass third companion (P$_{\mathrm{long}}$ = 99.3 days), of
spectral type mid-K or possibly even a white dwarf.  The close binary system showed eclipses from the earliest reported observations in 1899
until about 1918, when the eclipses stopped; eclipses reappeared in 1963 and were observed until about 1986.  Interpreting the
appearance of eclipses due to nodal regression, \cite{1999AJ....117..541L} derive a nodal period of 68 years for V~907~Sco.  Using Eq.~\ref{omegadotsimple} and assuming a mass of 2 M$_{\odot}$ for the host ($m_o$) and a mass of 0.5 M$_{\odot}$ for the companion
($m_p$), we compute a nodal regression period of 47.6~years, which agrees well with the nodal period
estimated by  \cite{1999AJ....117..541L}.  We therefore conclude that Eq.~\ref{omegadotsimple} also provides a reasonable description of the transient
eclipse observations of V~907~Sco.

\subsubsection{Application to TrES 2}

Applying now Eq.~\ref{omegadotsimple} to the TrES-2 exoplanet we express the period of the (unknown) perturbing planet
as a function of its also unknown mass through

\begin{equation}
P_p = \sqrt{\frac{m_p} {m_p + m_o} \frac{3 \pi P} {2} \frac{1} {\mathrm{d} \Omega/\mathrm{d}t}}. 
\label{tres2}
\end{equation}

Since the host star mass and the nodal regression are known, the perturbing mass is the only remaining unknown;
we note that Eq.~\ref{tres2} should be correct, as long as there is only one dominant perturber in the system with a low-eccentricity
orbit sufficiently far away from the known close-in transiting planet.  In Fig.~\ref{fig9} we plot the expected perturber
period $P_p$ as a function of $m_p$ (in Jupiter masses) for $m_o$ = 1~M$_{\odot}$ and the measured rate of nodal regression assuming
a linear coefficient of unity in Eq.~\ref{adot}.
Assuming {\it ad hoc} a mass of about one Jupiter mass for this perturber and taking into account that a factor of a few is likely (cf., Fig.~\ref{fig8}), we find that periods of 50 to 100~days are required
to explain the observed nodal regression in TrES-2.  Such an additional planet should be relatively easily detectable with RV studies.
\cite{2009A&A...498..567D} report the presence of a faint companion about one arcsec away from TrES-2.  Assuming this companion to be physical, 
a spectral type between K4.5 and K6, a mass of about 0.67~M$_{\odot}$ at a distance of 230~AU with a period of 3900~years follow.
Computing with these numbers the maximally expected nodal regression from Eq.~\ref{omegadotsimple}, one finds values a couple of orders of magnitude below the observed
values.  We thus conclude that this object cannot be the cause of the observed orbit variations.  On the other
hand, this companion, again if physical,  makes TrES-2 particularly interesting because it provides another cases of a planet/planetary system 
in a binary system, and eventually the orbit planes of binary system and the planet(s) can be derived.

\begin{figure}[ht]
\centering
\includegraphics[width=8.0cm]{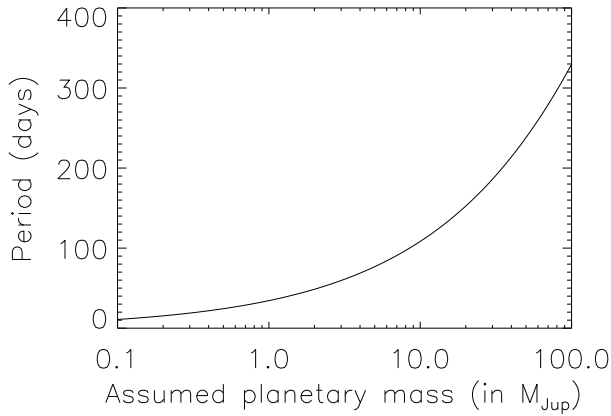}
\caption{Period of hypothesized second planet vs. mass assuming a linear coefficient of unity in Eq.~\ref{adot} and near coplanarity.}
\label{fig9}
\end{figure}

\subsection{Transit timing variations by a putative perturber}

A perturbing second planet capable of causing fast nodal precession on the
transiting planet is also expected to induce short-term periodic variations of its orbital elements.  In addition to the
secular precession of the node of the orbit we would thus expect to see short-term transit timing variations (TTVs), periodic variations of the mid-transit times (\cite{2005Sci...307.1288H, 2005MNRAS.359..567A}). Just as nodal regression,  the TTV signal can be used to find and characterize planetary companions of transiting exoplanets. \cite{2009arXiv0909.1564R} carefully analyzed eight transit light curves of TrES-2 over several years. However, they were unable to detect any statistically significant TTV amplitudes in the TrES-2b light curves above 
about 50~s; cf.~their Fig.~10.
% Their results are hampered by the large error of several tenths of seconds. 
Therefore, the existence of perturbing objects leading to TTVs on the scale of up to about 50~s are consistent with actual observations. Putting it differently, the orbital parameters of any perturbing object causing the nodal precession of the orbit should yield a TTV amplitude below that and hence remain undetactable in the presently existing data.

To analyze the mutual gravitational influence of a perturbing second planet in the system on TrES-2, we have to treat the classical three-body problem of celestial mechanics. Instead of direct $N$-body integrations of the equations of motion we use an alternative method based on analytic perturbation theory developed and extensively tested by \cite{2008ApJ...688..636N} and \cite{2009ApJ...701.1116N}. Outside possible mean-motion resonances their approach allows for a fast computation of the expected TTV amplitude given a combination of system parameters. As input we have to specify the orbital elements and masses of both planets. Consistent with the observations
we assume TrES-2 to be in a circular orbit around its host star, while we allow for different eccentricities $e_p$ and periods $P_p$ of the perturber, which we assume to be of Jovian mass; since the TTV amplitudes scale nearly linearly with the perturber mass, we confine our treatment to $m_p=1\ M_{\mathrm{Jup}}$; all other orbital elements are set to zero. This is justified as these parameters in most cases do not lead to a significant amplification of the TTV signal (see \cite{2009ApJ...701.1116N} for a detailed discussion of the impact of these orbital elements). %We tried other combinations but found no significantly different results. 
The resulting TTV amplitude for different reasonable orbit configurations (given the observed secular node regression as discussed above)
of the system of P$_{pert}$ = 30, 50 and 70~days is plotted vs. the assumed eccentricity in Fig.~\ref{ttvovere2}; the currently
available upper limit to any TTV signal derived by \cite{2009arXiv0909.1564R} is also shown.  As is obvious from
Fig.~\ref{ttvovere2}, a Jovian-mass perturber at a distance required to impose the observed secular changes (period of $50-100$~days) leads to a TTV signal well below the current detection limit for all eccentricities $e_p$ as long as $e_p \lesssim 0.4$. 
%Only an unusually high eccentricity of the perturber could cause an observable signal. 
We therefore conclude that a putative perturbing Jovian-mass planet with a moderate eccentricity and 
with a period between $30-70$~days would not yield any currently detectable TTV signal and would therefore be a valid
explanation for the observed inclination change in the TrES-2 system. 

\begin{figure}[ht]
\centering
\includegraphics[width=8.0cm]{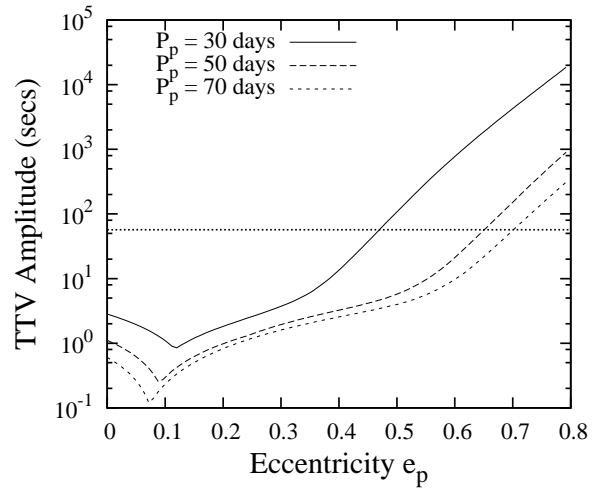}
\caption{Amplitude of expected Transit Timing Variations (TTVs) in the TrES-2 system. The perturber is assumed to have $m_p=1\ M_{\mathrm{Jup}}$. Its eccentricity $e_p$ and period $P_p$ are varied within plausible ranges. The orbit of TrES-2 is assumed to be circular. The vertical dotted line marks the best fitting TTV signal found by \cite{2009arXiv0909.1564R} of 57~s.}
\label{ttvovere2}
\end{figure}

\section{Conclusions}

%%%%%%%%%%%%%%%%%%%%%%%%%%%%%%%%%%%%%%%
% Observational part
%%%%%%%%%%%%%%%%%%%%%%%%%%%%%%%%%%%%%%%

In summary, our new observations taken in the spring of 2009 confirm the smaller transit durations reported in Paper I
and suggest an even further decrease.  With our simultaneously taken multicolor BUSCA data  we demonstrate that the recorded multicolor lightcurves can be consistently modelled with a reasonable set of limb darkening coefficients, and that there is no need to 
fit the limb darkening coefficients to any particular light curve.  An error in the description of the limb darkening therefore appears
thus as an unlikely cause of the observed inclination changes.   
Also as expected, the obtained stellar radius is independent of the wavelength band used, demonstrating the internal
self-consistency of our modelling. 

%Our new transit observations of the TrES-2 exoplanet taken in 
%the spring of 2009 confirm the smaller transit durations reported in Paper I.  Our fit results suggest that the duration has even changed since September 2008.  For the OLT observations in September 2008 
%and April 2009 we actually used exactly the same setup, viz., same telescope, the same filter (I-filter) and the same model analysis. 
%In Paper I we noted an apparent decrease of the orbit inclination of TrES-2 and predicted inclination values below the first transit threshold ($i_{min,1}<83.417^{\circ}$) after October 2008.  Both our new 
%OLT data set and all BUSCA channel observations provide further indications for this effect: All five independent measurements all yield only downward excursions (Fig. \ref{fig7}). We consider it unlikely that such an 
%systematic trend could be produced by accident.

%%%%%%%%%%%%%%%%%%%%%%%%%%%%%%%%%%%%%%%
% Theoretical part
%%%%%%%%%%%%%%%%%%%%%%%%%%%%%%%%%%%%%%%

As to the possible causes for the observed apparent orbit inclination change in TrES-2 we argue that the apparent observed inclination change is very likely caused by nodal regression.  The assumption of an oblate host star leads to implausibly 
large $J_2$ coefficients, we therefore favor an explanation with a third body.  We argue that Eq.~\ref{omegadotsimple} is a reasonable approximation for the interpration of the observed inclination changes; applying it to the TrES-2 system, we find that a planet of
one Jovian mass with periods between $50-100$ days would suffice to cause the observed inclination changes, while at the same time
yield TTVs with amplitudes well below the currently available upper limits.

%Interpreting this nodal regression as a consequence of an oblate host star, requires extremely large $J_2$ coefficients, which are quite implausible, given that the TrES-2 host star is a slow rotator.
%Furthermore, observations of the Rossiter-McLaughlin effect for close-in
%planets usually show very good alignment between the rotation axes of the host stars and orbital planes of their close-in planets,
%rendering the orbit regression rather difficult to observe.  
%Thus, an oblate host star is not a very likely cause for the observed inclination change.
%We next analyze the perturbations caused by an hypothetical third body; since such a third body is unknown at present, it seems reasonable to consider only three-body (rather than $N$-body) perturbations.  Based on our solar system and the apparently very similar case of the transient triple existing binary V~907~Sco we
%argue that Eq.\ref{omegadotsimple} is a good approximation for the TrES-2 system and we use it to relate mass and period of this hypothesized third body. Based on this analysis the distant companion around TrES-2 detected by \cite{2009A&A...498..567D} cannot be held responsible for the observed orbit variations; if one assumes about a Jupiter mass for this hypothesized new planet, an orbital period of about 50 -- 100 days results. 
The assumption of such an additional planet in the TrES-2 system is 
entirely plausible.  First of all, if it is near coplanar with TrES-2b, it would not cause any eclipses and therefore remain undetected in transit searches.  Next, an inspection of the exosolar planet data base maintained at {\it www.exoplanet.eu} reveals a number of exoplanet systems with properties similar to those postulated for TrES-2, i.e., a close-in planet together with a massive planet further out:  In the Gl~581 system there is a 0.02~Jupiter-mass planet with a period of 66~days, and in fact a couple of similarly massive planets further in with periods of 3.1, 5.4 and 12.9~days 
respectively;  in the system HIP~14810 there is a close-in planet with a 6.6~day period and a somewhat lighter planet with a period of 147~days, in the HD~160691 system the close-in planet has a period of 9.6~days and two outer planets with Jupiter masses are known with periods of 310 and 643~days.  It is also clear that in these systems nodal regression changes must occur, unless these systems are exactly coplanar, which appears unlikely.  Therefore on longer time scales the observed orbit inclination in these systems must change, but only in transiting systems the orbit inclination can be measured
with sufficient accuracy. 
%We showed that such an additional planet of Jovian mass would be entirely consistent with the current lack of observational evidence for short-term TTVs as the expected TTV amplitudes are usually around 10~s. Vice versa, the TrES-2 system with one or possibly more planets further out would be by no means unusual, and it should be rather straightforward to detect the required planet(s) from RV monitoring. 
Because of its apparent inclination change TrES-2b is clearly among the more interesting extrasolar planets. If the system continues its behavior in the future the transits of TrES-2b will disappear.  Fortunately, 
%As shown above, such features can be naturally explained by the presence of a third body in the form of another planet. It would clearly be interesting to acquire accurate spectroscopic data and, further, we 
within the first data set of the  Kepler mission $\sim$~30 transits should be covered. From our derived inclination change rate of $\Delta i \sim 0.075^{\circ}/\textrm{yr}$ this corresponds to an overall change of $\Delta i \sim 0.015^{\circ}$ in this first data set, which ought to be detectable given the superior accuracy of the space-based Kepler photometry.  As far as the detection of our putative second planet is concerned, RV methods appear to be more promising than a search for TTVs, unless the orbital eccentricities are very large.

\begin{acknowledgements}

The paper is based on observations collected with the 2.2 m telescope at the Centro Astron\'{o}mico Hispano Alem\'{a}n (CAHA) at Calar Alto (Almer\'{i}a, Spain), operated jointly by the Max-Planck 
Institut f\"{u}r Astronomie and the Instituto de Astrof\'{i}sica de Andaluc\'{i}a (CSIC). The authors thank H.Poschmann and the BUSCA team for their tremendous work on the new controller and CCD system. 
DM was supported in the framework of the DFG-funded Research Training Group ``Extrasolar Planets and their Host Stars'' (DFG 1351/1). SS acknowledges DLR support through the grant 50OR0703.

\end{acknowledgements}

\bibliographystyle{aa}

\bibliography{aa}

\end{document}